\documentclass[11pt]{revtex4}    
\usepackage{amssymb,epsf}
\usepackage{latexsym}
\begin{document}

\title{ Ricci Flat Rotating Black Branes with a Conformally Invariant
Maxwell Source}
\author{S. H. Hendi$^{1,2}$\footnote{hendi@mail.yu.ac.ir} and H. R. Rastegar-Sedehi$^{3}$}

\address{$^1$ Physics Department,
College of Sciences, Yasouj University, Yasouj
75914, Iran\\
$^2$ Research Institute for Astrophysics and Astronomy of Maragha
(RIAAM), P.O. Box 55134-441, Maragha, Iran\\
$^3$Physics Department and Biruni Observatory, College of
Sciences, Shiraz University, Shiraz 71454, Iran}

\begin{abstract}
We consider Einstein gravity coupled to an $U(1)$ gauge field for which the
density is given by a power of the Maxwell Lagrangian. In $d$-dimensions the
action of Maxwell field is shown to enjoy the conformal invariance if the
power is chosen as $d/4$. We present a class of charge rotating solutions in
Einstein-conformally invariant Maxwell gravity in the presence of a
cosmological constant. These solutions may be interpreted as black brane
solutions with inner and outer event horizons or an extreme black brane
depending on the value of the mass parameter. Since we are considering power
of the Maxwell density, the black brane solutions exist only for dimensions
which are multiples of four. We compute conserved and thermodynamics
quantities of the black brane solutions and show that the expression of the
electric field does not depend on the dimension. Also, we obtain a
Smarr-type formula and show that these conserved and thermodynamic
quantities of black branes satisfy the first law of thermodynamics. Finally,
we study the phase behavior of the rotating black branes and show that there
is no Hawking--Page phase transition in spite of conformally invariant
Maxwell field.
\end{abstract}

\maketitle

\address{$^1$ Physics Department, College of Sciences, Yasouj University, Yasouj 75914,
Iran \\
         $^2$ Research Institute for Astronomy and Astrophysics of Maragha (RIAAM)- Maragha, P.O. Box: 55134-441, Iran\\
         $^3$ Physics Department and Biruni Observatory, College of Sciences, Shiraz University, Shiraz 71454, Iran}

\address{$^1$ Physics Department, College of Sciences, Yasouj University,
Yasouj 75914, Iran \\
$^2$ Research Institute for Astrophysics and Astronomy of Maragha
(RIAAM), Maragha 55134, Iran \\
$^3$Physics Department and Biruni Observatory, College of
Sciences, Shiraz University, Shiraz 71454, Iran}

\section{Introduction}

The conjectured equivalence of string theory on anti-de Sitter (AdS) spaces
(times some compact manifold) and certain superconformal gauge theories
living on the boundary of AdS \cite{Mal} has lead to an increasing interest
in asymptotically anti-de Sitter black holes. This conjecture is now a
fundamental concept that furnishes a means for calculating the action and
conserved quantities intrinsically without reliance on any reference
spacetime \cite{Sken}. Also, coupling between gravity and gauge fields is a
common feature of many unification theories in higher dimensions \cite%
{Bailin}. In supergravity theories, such gauge fields are frequently
essential in order to complete the multiple structure and to guarantee the
invariance of the Lagrangian with respect to local supersymmetry
transformations \cite{Nieu}. They can provide, in some cases, a dynamical
mechanism for the compactification of extra dimensions \cite{Freu}.

In this supersymmetric context, they can lead to a geometric interpretation
in the superspace. The simplest example of coupling of a gauge field to
gravity is the Einstein-Maxwell system in four dimensions, whose extension
to higher dimensions can lead to some interesting features: its reduction to
four dimensions implies a non-trivial coupling of the electromagnetic field
to gravity. In the conventional, straightforward generalization of the
Maxwell field to higher dimensions one essential property of the
electromagnetic field is lost, namely, conformal invariance.

The Reissner-Nordstr\"{o}m black hole, the first solution for which the
matter source is conformally invariant, is an electrically charged but
non-rotating black hole solution in four dimensions. There exists a
conformally invariant extension of the Maxwell (CIM) action in higher
dimensions, if one uses the lagrangian of the $U(1)$ gauge field in the form:%
\begin{equation}
I_{CIM}=\alpha \int d^{d}x\sqrt{-g}\left( F_{\mu \nu }F^{\mu \nu }\right)
^{d/4}  \label{CIM}
\end{equation}%
where $F_{\mu \nu }=\partial _{\mu }A_{\nu }-\partial _{\nu }A_{\mu }$ is
the Maxwell tensor and $\alpha $ is a constant. It is straightforward to
show that the action (\ref{CIM}) is invariant under conformal transformation
($g_{\mu \nu }\longrightarrow \Omega ^{2}g_{\mu \nu }$ and $A_{\mu
}\longrightarrow A_{\mu }$) and for $d=4$, the action (\ref{CIM}) reduces to
the Maxwell action as it should be. The energy-momentum tensor associated to
$I_{CIM}$ is given by
\begin{equation}
T_{\mu \nu }=\alpha \left( dF_{\mu \rho }F_{\nu }^{~\rho }F^{(d/4)-1}-g_{\mu
\nu }F^{d/4}\right)   \label{T}
\end{equation}%
where $F=F_{\mu \nu }F^{\mu \nu }$ and it is easy to show that $T_{\mu
}^{\mu }=0$ \cite{comment}. As one can see the clue of the conformal
invariance lies in the fact that we have considered power of the Maxwell
invariant. This idea has been applied in the case of scalar field for which
it has been shown that particular power of the massless Klein-Gordon
Lagrangian exhibits conformal invariance in arbitrary dimension \cite%
{Haissene}. It would be interesting to see whether black hole solutions can
also be obtained in this case.

In what follows, we have been considered the action (\ref{CIM}) as the
matter source coupled to the Einstein gravity to take advantage of the
conformal symmetry to construct the analogues of the four-dimensional
Reissner-Nordstr\"{o}m black hole solutions in higher dimensions. The form
of the energy-momentum tensor (\ref{T}), automatically restricts the
dimensions to be only multiples of four.

However, the black hole solutions which have been studied so far represent
only a special case of the general black hole solution since rotation has
been ignored \cite{Freu,AllStaticEM}. To see whether qualitative properties
of generic black holes depend on the matter content, one must consider
rotating charged black holes \cite{AllRotBH,dehhend2,Rastegar}. This is the
subject of the present investigation. Indeed, in this paper we obtain a
class of higher dimensional solution of rotating black branes with a
conformally invariant Maxwell source in the presence of cosmological
constant. These solutions contain three conserved quantities, the mass $M$,
electrical charge $Q$, and angular momentum $J$. By calculation of other
thermodynamic quantities, we show that thermodynamic and conserved
quantities satisfy the first law of thermodynamics. Also, we investigate the
stability of charged rotating black brane solutions in the canonical and
grand canonical ensembles. In fact, the Hawking-Page phase transition for
the Schwarzschild-AdS black hole \cite{Hawpag} does not occur for
asymptotically locally anti-de Sitter black holes whose horizons have
vanishing or negative constant curvature, and they are thermodynamically
--locally-- stable since they have a positive heat capacity \cite{Birmingham}%
. The inclusion of electric charge further modifies the thermodynamical
properties of black holes and a more complex phase structure arises,
allowing for an analogous of the Hawking-Page phase transition \cite{BCCC}.

The rest of the paper is organized as follows. In Sec. \ref{Fiel},
we give a brief review of the field equations of Einstein gravity
with conformally invariant Maxwell field in the presence of
cosmological constant and review the counterterm method for
calculating the conserved quantities of the solutions. In Sec.
\ref{Static sol}, we obtain the $(4p+4)$-dimensional static
solutions of field equations, which are asymptotically (anti)-de
sitter or flat. Then, we present a new class of charged rotating
black brane solutions and investigate their properties in Sec.
\ref{Sol}. In Sec. \ref{Therm} we obtain mass, angular momentum,
entropy, temperature, charge, and electric potential of the
rotating black brane solutions and show that these quantities
satisfy the first law of thermodynamics. We also perform a local
stability analysis of the black branes in the canonical and
grand-canonical ensembles. We finish our paper with some
concluding remarks.

\section{Field Equations in Einstein-Conformally Invariant Maxwell Gravity
\label{Fiel}}

In $d=4$, the action of Einstein-Maxwell is trivial, but for $d={4p+4}$
dimensions with ${p\in \mathbf{N}}$, we consider the Einstein action with a
conformally invariant Maxwell action given by
\begin{eqnarray}
I_{G}=&-&\frac{1}{16\pi }\int_{\mathcal{M}}d^{4p+4}x\sqrt{-g}\left[ \mathcal{%
R}-2\Lambda -\alpha (F^{\mu \nu }F_{\mu \nu })^{p+1}\right] \\
&-&\frac{1}{8\pi }\int_{\partial \mathcal{M}}d^{4p+3}x\sqrt{_{-}\gamma }%
\Theta (\gamma ),  \nonumber  \label{action}
\end{eqnarray}
where $\Lambda $ is the negative cosmological constant which is equal to $%
-(4p+3)(2p+1)/l^{2}$ for asymptotically AdS solutions, where $l$\ is a scale
length factor; $\alpha $ is a constants and $\mathcal{R}$ is scalar
curvature. The second integral in Eq. (\ref{action}) is a boundary term
which is chosen such that the variational principle is well defined \cite%
{Myers}.\ In this term\ $\gamma _{ij}$\ is an induced metric on the boundary
$\partial \mathcal{M}$ and $\Theta $ is the trace of the extrinsic curvature
$\Theta _{\mu \nu }$ of any boundary of the manifold $\mathcal{M}$. Varying
the action with respect to the gauge field $A_{\mu }$ and the metric $g_{\mu
\nu }$, the field equations are obtained as
\begin{equation}
\partial _{\mu }\left( \sqrt{-g}F^{\mu \nu }F^{p}\right) =0,  \label{eqME1}
\end{equation}
\begin{equation}
G_{\mu \nu }+\Lambda g_{\mu \nu }=2\alpha \left[ (p+1)F_{\mu \rho }F_{\nu
}^{~\rho }F^{p}-\frac{1}{4}g_{\mu \nu }F^{p+1}\right] ,  \label{eqME2}
\end{equation}
respectively.

In general the action $I_{G}$, is divergent when evaluated on the solutions,
as is the Hamiltonian and other associated conserved quantities. A
systematic method of dealing with this divergence in Einstein gravity is
through the use of the counterterms method inspired by the anti-de Sitter
conformal field theory (AdS/CFT) correspondence \cite{Mal}. In Einstein
gravity, there exists only one counterterm and therefore the total finite
action is
\begin{equation}
I=I_{G}+I_{\mathrm{ct}},  \label{Itot}
\end{equation}%
where the counterterm action is
\begin{equation}
I_{\mathrm{ct}}=-\frac{2p+1}{4\pi l}\int_{\partial \mathcal{M}}d^{4p+3}x%
\sqrt{-\gamma }.  \label{Ict}
\end{equation}%
Having the total finite action one can construct a divergence-free stress
energy tensor by use the Brown-York definition \cite{Brown}. One can show
that the finite stress energy tensor is
\begin{equation}
T^{ab}=\frac{1}{8\pi }\left[ \Theta ^{ab}-\Theta \gamma ^{ab}-\frac{2(2p+1)}{%
l}\gamma ^{ab}\right] .  \label{Stres}
\end{equation}%
To compute the conserved charges of the spacetime, we choose a spacelike
hypersurface $\mathcal{B}$ in $\partial \mathcal{M}$ with metric $\sigma
_{ij}$, and write the boundary metric in \textit{ADM}
(Arnowitt-Deser-Misner) form
\begin{equation}
\gamma _{ab}dx^{a}dx^{a}=-N^{2}dt^{2}+\sigma _{ij}\left( d\varphi
^{i}+V^{i}dt\right) \left( d\varphi ^{j}+V^{j}dt\right) ,
\end{equation}%
where the coordinates $\varphi ^{i}$ are the angular variables
parameterizing the hypersurface of constant $r$ around the origin, and $N$
and $V^{i}$ are the lapse and shift functions respectively. When there is a
Killing vector field $\mathcal{\xi }$ on the boundary, then the quasilocal
conserved quantities associated with the stress energy tensors of Eq. (\ref%
{Stres}) can be written as
\begin{equation}
\mathcal{Q}(\mathcal{\xi )}=\int_{\mathcal{B}}d^{4p+2}\varphi \sqrt{\sigma }%
T_{ab}n^{a}\mathcal{\xi }^{b},  \label{charge}
\end{equation}%
where $\sigma $ is the determinant of the metric $\sigma _{ij}$, and $n^{a}$
is the timelike unit normal vector to the boundary $\mathcal{B}$. Hereafter
we set $\alpha =(-1)^{p}$\ without loss of generality and consequently the
energy density (the $T_{\widehat{0}\widehat{0}}$ component of the
energy-momentum tensor in the orthonormal frame) is positive.

\section{The Static Solutions \label{Static sol}}

Here we want to obtain the $(4p+4)$-dimensional static solutions of Eq. (\ref%
{eqME1}) and (\ref{eqME2}), which are asymptotically (anti)-de sitter or
flat. We assume that the metric has the following form:
\begin{equation}
ds^{2}=-f(r)dt^{2}+\frac{dr^{2}}{f(r)}+r^{2}d\Omega _{4p+2}^{2}  \label{met}
\end{equation}%
where $d\Omega _{4p+2}^{2}$ is
\[
d\Omega _{4p+2}^{2}=\left\{
\begin{array}{cc}
d\theta _{1}^{2}+\sum\limits_{i=2}^{4p+2}\prod\limits_{j=1}^{i-1}\sin
^{2}\theta _{j}d\theta _{i}^{2} & k=1 \\
d\theta _{1}^{2}+\sinh ^{2}\theta _{1}d\theta _{2}^{2}+\sinh ^{2}\theta
_{1}\sum\limits_{i=3}^{4p+2}\prod\limits_{j=2}^{i-1}\sin ^{2}\theta
_{j}d\theta _{i}^{2} & k=-1 \\
\sum\limits_{i=1}^{4p+2}d\zeta _{i}^{2} & k=0%
\end{array}%
\right.
\]%
which represents the line element of an $(4p+2)$-dimensional hypersurface
with constant curvature $(4p+2)(4p+1)k$ and volume $V_{4p+2}$. We use the
gauge potential ansatz
\begin{equation}
A_{\mu }=h(r)\delta _{\mu }^{0}
\end{equation}%
in conformally invariant Maxwell equation (\ref{eqME1}). we obtain
\[
h(r)=\frac{-q}{r},
\]%
where $q$ is an integration constant which is related to the electric charge
parameter. Also, the conformally invariant Maxwell equation implies that the
electric field in $(4p+4)$-dimensions is given by%
\begin{equation}
F_{tr}=\frac{q}{r^{2}}.
\end{equation}%
It is interesting to note that the expression of the electric field does not
depend on the dimension and its value coincides with the Reissner-Nordstr%
\"{o}m solution in four dimensions.

To find the metric function $f(r)$, one may use any components of Eq. (\ref%
{eqME2}). The simplest equation is the $rr$ component of these equations,
which can be written as
\begin{equation}
f^{^{\prime }}(r)+\frac{4p+1}{r}\left[ f(r)+k\right] +\frac{2^{p}q^{2p+2}}{%
r^{4p+3}}+\frac{\Lambda r}{2p+1}=0,  \label{rrcomp}
\end{equation}%
where the prime denotes a derivative with respect to $r$. The solutions of
Eq. (\ref{rrcomp}) can be written as
\begin{equation}
f(r)=k+\frac{r^{2}}{l^{2}}-\frac{m}{r^{4p+1}}+\frac{2^{p}q^{2p+2}}{r^{4p+2}},
\label{f}
\end{equation}%
where $m$ is other integration constant proportional to the mass parameter.
One can check that the solution given by Eq. (\ref{f}) satisfies all the
components of the field equations (\ref{eqME2}). The metric function $f(r)$,
presented here, differ from the standard higher-dimensional solutions; we
find that the electric charge term in the metric function goes as $%
r^{-(4p+2)}$ and in the standard case is $r^{-2(4p+1)}$. Hereafter we set $%
k=0$ and investigate the proprties of the boundary flat black brane
solutions.

\section{The $(4p+4)$-dimensional Charged Rotating Black Branes\label{Sol}}

The metric of $(4p+4)$-dimensional asymptotically AdS charged rotating black
brane with $\kappa $ rotation parameters and flat boundary at constant $t$
and $r$ may be written as \cite{LemAw}
\begin{eqnarray}
ds^{2} &=&-f(r)\left( \Xi dt-{{\sum_{i=1}^{\kappa }}}a_{i}d\phi _{i}\right)
^{2}+\frac{r^{2}}{l^{4}}{{\sum_{i=1}^{\kappa }}}\left( a_{i}dt-\Xi
l^{2}d\phi _{i}\right) ^{2}  \nonumber \\
&&\;+\frac{dr^{2}}{f(r)}-\frac{r^{2}}{l^{2}}{\sum_{i<j}^{\kappa }}%
(a_{i}d\phi _{j}-a_{j}d\phi _{i})^{2}+r^{2}dX^{2},  \label{met2}
\end{eqnarray}%
where $f(r)$ is the same as $f(r)$ given in Eq. (\ref{f}) with $k=0$, $\Xi =%
\sqrt{1+\sum_{i}^{k}a_{i}^{2}/l^{2}}$ and $dX^{2}$ is the Euclidean metric
on the $\left( 4p+2-\kappa \right) $-dimensional submanifold. The rotation
group in $d$-dimensions is $SO(4p+3)$ and therefore the maximum number of
rotation parameters in $(4p+4)$-dimensions is $[\frac{4p+3}{2}]$, where $[z]$
denotes the integer part of $z$.

The gauge potential is%
\begin{equation}
A_{\mu }=\frac{-q}{r}\left( \Xi \delta _{\mu }^{0}-\delta _{\mu
}^{i}a_{i}\right) \hspace{0.5cm}(\mathrm{no\;sum\;on}\;i).
\end{equation}

\subsection{Properties of the solutions}

In order to study the general structure of these solutions, we first look
for the essential singularities. After some algebraic manipulation, one can
show that the Kretschmann scalar in ($4p+4$)-dimensions,
\begin{eqnarray}
R_{\mu \nu \rho \sigma }R^{\mu \nu \rho \sigma } &=&\frac{8(p+1)(4p+3)}{l^{4}%
}+\frac{4(4p+3)(4p+1)(2p+1)^{2}m^{2}}{r^{8p+6}}-  \nonumber \\
&&\frac{2^{p+4}(p+1)(4p+3)(2p+1)(4p+1)mq^{2p+2}}{r^{8p+7}}+  \nonumber \\
&&\frac{2^{2p+3}(p+1)(2p+1)(16p^{2}+24p+7)q^{4p+4}}{r^{8p+8}},
\label{kretch}
\end{eqnarray}
diverges at $r=0$ and is finite for $r\neq 0$. Thus, there is an
curvature timelike singularity located at $r=0$. In the case of a
vanishing cosmological constant, the solutions given in Eqs.
(\ref{f}) and (\ref{met2}) are not well-behaved asymptotically and
for the case of a positive cosmological constant [$\Lambda
=(4p+3)(2p+1)/l^{2}$] the solutions are asymptotically dS, but the
singularity is naked (see the Appendix of \cite{dehhend2}).

It is proved that a stationary black hole event horizon should be a Killing
horizon in the four-dimensional Einstein gravity \cite{Haw1}. In our
solutions the Killing vector,
\begin{equation}
\chi =\partial _{t}+{\sum_{i}^{k}}\Omega _{i}\partial _{\phi _{i}},
\label{Kil}
\end{equation}%
is the null generator of the event horizon. The angular velocities $\Omega
_{i}$ are \cite{dehbordshah}
\begin{equation}
\Omega _{i}=\frac{a_{i}}{\Xi l^{2}}.  \label{Om}
\end{equation}%
The temperature may be obtained through the use of the definition of surface
gravity,
\[
T_{+}=\frac{1}{\beta _{+}}=\frac{1}{2\pi }\sqrt{-\frac{1}{2}\left( \nabla
_{\mu }\chi _{\nu }\right) \left( \nabla ^{\mu }\chi ^{\nu }\right) }
\]%
where $\chi $ is the Killing vector (\ref{Kil}). One obtains
\begin{equation}
T_{+}=\frac{(4p+3)r_{+}^{4p+4}-2^{p}q^{2p+2}l^{2}}{4\pi l^{2}\Xi r_{+}^{4p+3}%
}.  \label{Temp}
\end{equation}%
The event horizon of solutions are located at the root(s) of $g^{rr}=f(r)=0$%
. Indeed, the metric of Eqs. (\ref{f}) and (\ref{met2}) has two inner and
outer event horizons located at $r_{-}$ and $r_{+}$, provided the mass
parameter $m$ is greater than $m_{\mathrm{ext}}$ given as
\begin{equation}
m_{\mathrm{ext}}=\frac{4\left( p+1\right) }{l^{2}}\left( \frac{2^{p}l^{2}q_{%
\mathrm{ext}}^{2p+2}}{4p+3}\right) ^{\left( 4p+3\right) /\left( 4p+4\right)
}.  \label{mext}
\end{equation}%
The solutions present naked singularity for $m<m_{\mathrm{ext}}$ and when $%
m=m_{\mathrm{ext}}$, we have an extreme black brane with horizon radius
\begin{equation}
r_{+\mathrm{ext}}=\left( \frac{2^{p}l^{2}q_{\mathrm{ext}}^{2p+2}}{4p+3}%
\right) ^{1/(4p+4)}.  \label{rext}
\end{equation}

\section{Thermodynamics \label{Therm}}

\subsection{Conserved and thermodynamic quantities}

Usually entropy of the black holes satisfies the so-called area law of
entropy which states that the black hole entropy equals to one-quarter of
horizon area \cite{Beck}. Since the area law of the entropy is universal,
and applies to all kinds of black holes/branes in Einstein gravity \cite%
{Beck}, therefore the entropy per unit volume $V_{4p+2}$ is equal to
one-quarter of the area of the horizon, i.e.,
\begin{equation}
S=\frac{\Xi }{4}r_{+}^{2(2p+1)}.  \label{Entropy}
\end{equation}
where $V_{4p+2}$ is the volume of the boundary at constant $t$ and $r$.

The electric charge per unit volume $V_{4p+2}$ of the black brane, $Q$, can
be found by calculating the flux of the electromagnetic field at infinity,
yielding
\begin{equation}
Q=\frac{2^{p}(p+1)\Xi q^{2p+1}}{4\pi }.  \label{Charg}
\end{equation}
Also, the electric potential $U$, measured at infinity with respect to the
horizon, is defined by \cite{Gub}
\begin{equation}
U=A_{\mu }\chi ^{\mu }\left\vert _{r\rightarrow \infty }-A_{\mu }\chi ^{\mu
}\right\vert _{r=r_{+}},  \label{Pot1}
\end{equation}
where $\chi $ is the null generator of the horizon given by Eq. (\ref{Kil}).
One finds
\begin{equation}
U=\frac{q}{\Xi r_{+}},  \label{Pot2}
\end{equation}
which is independent of dimensions.

The present spacetime (\ref{met2}), have boundaries with timelike ($\xi
=\partial /\partial t$) and rotational ($\varsigma =\partial /\partial
\varphi $) Killing vector fields. From Eq. (\ref{charge}), one obtains the
quasilocal mass and angular momentum
\begin{eqnarray}
M &=&\int_{\mathcal{B}}d^{4p+2}\varphi \sqrt{\sigma }T_{ab}n^{a}\xi ^{b},
\label{Mas} \\
J &=&\int_{\mathcal{B}}d^{4p+2}\varphi \sqrt{\sigma }T_{ab}n^{a}\varsigma
^{b},  \label{Amom}
\end{eqnarray}%
provided the hypersurface $\mathcal{B}$ contains the orbits of $\varsigma $.
Using the Eqs. (\ref{Mas}) and (\ref{Amom}), we find the total mass per unit
volume $V_{4p+2}$ of the solutions to be given by
\begin{equation}
M=\frac{\left[ \left( 4p+3\right) \Xi ^{2}-1\right] }{16\pi }m,  \label{Mass}
\end{equation}%
while the angular momenta per unit volume $V_{4p+2}$ are given by
\begin{equation}
J_{i}=\frac{\left( 4p+3\right) \Xi m}{16\pi }a_{i}.  \label{Angmom}
\end{equation}

\subsection{Energy as a function of entropy, angular momenta and charge}

Now, we check the first law of thermodynamics for our solutions in
Einstein-conformally invariant Maxwell gravity. In order to do this, we
obtain the mass $M$ as a function of the extensive quantities $S$, $\mathbf{J%
}$ and $Q$. Using the expression for the entropy, the mass, the angular
momenta, and the charge given in Eqs. (\ref{Mass}), (\ref{Angmom}), (\ref%
{Entropy}), (\ref{Charg}), and the fact that $f(r_{+})=0$, one can obtain a
Smarr-type formula as
\begin{equation}
M(S,J,Q)=\frac{\left[ \left( 4p+3\right) Z-1\right] J}{\left( 4p+3\right) l%
\sqrt{Z(Z-1)}},  \label{Smar}
\end{equation}%
where $J=\left\vert \mathbf{J}\right\vert =\sqrt{\sum_{i}^{k}J_{i}^{2}}$ and
$Z=\Xi ^{2}$ is the positive real root of the following equation
\begin{equation}
S^{(4p+3)/(4p+2)}+\frac{\pi Ql^{2}}{p+1}\left( \frac{\pi ^{2}Q^{2}}{%
2^{2p}\left( p+1\right) ^{2}S}\right) ^{1/(4p+2)}-\frac{2^{2p/(2p+1)}\pi
lJZ^{1/(8p+4)}}{(4p+3)\sqrt{Z-1}}=0.  \label{Zsmar}
\end{equation}%
One may then regard the parameters $S$, $\mathbf{J}$ and $Q$ as a complete
set of extensive parameters for the mass $M(S,\mathbf{J},Q)$ and define the
intensive parameters conjugate to $S$, $\mathbf{J}$ and $Q$. These
quantities are the temperature, the angular velocities and the electric
potential. It is a matter of straightforward calculations to obtain
\begin{equation}
\Omega _{i}=\left( \frac{\partial M}{\partial J_{i}}\right) _{S,Q},\ \
T=\left( \frac{\partial M}{\partial S}\right) _{J,Q},\ \ U=\left( \frac{%
\partial M}{\partial Q}\right) _{S,J}.  \label{Dsmar1}
\end{equation}%
Using equations (\ref{Entropy}), (\ref{Charg}) and (\ref{Angmom}), it is
easy to show that the intensive quantities calculated from equation (\ref%
{Dsmar1}) coincide with equations (\ref{Om}), (\ref{Temp}) and (\ref{Pot2}).
Thus, these thermodynamics quantities satisfy the first law of
thermodynamics
\begin{equation}
dM=TdS+{{{\sum_{i=1}^{k}}}}\Omega _{i}dJ_{i}+UdQ.  \label{Flth}
\end{equation}

\subsection{Stability in the canonical and the grand-canonical ensemble}

Finally, we investigate the stability of charged rotating black brane
solutions of Einstein-conformally invariant Maxwell gravity. The local
stability can in principle be carried out by finding the determinant of the
Hessian matrix of $M(S,Q,\mathbf{J})$ with respect to its extensive
variables $X_{i}$, $\mathbf{H}_{X_{i}X_{j}}^{M}=[\partial ^{2}M/\partial
X_{i}\partial X_{j}]$ \cite{Gub}. In our case the mass is a function of the
entropy, the angular momenta and the charge. The number of thermodynamic
variables depends on the ensemble that is used. In the canonical ensemble,
the charge and the angular momenta are fixed parameters, and therefore the
positivity of the heat capacity $C_{\mathbf{J},Q}=T_{+}/(\partial
^{2}M/\partial S^{2})_{\mathbf{J},Q}$ is sufficient to ensure the local
stability. $(\partial ^{2}M/\partial S^{2})_{\mathbf{J},Q}$ at constant
charge and angular momenta is
\begin{equation}
\left( \frac{\partial ^{2}M}{\partial S^{2}}\right) _{_{\mathbf{J},Q}}=\frac{%
\left( 4p+3\right) \left( r_{+}^{4\left( p+1\right) }+2^{p}q^{2\left(
p+1\right) }l^{2}\right) ^{-1}}{\pi \Xi ^{2}l^{2}\left[ (p+1)\Xi ^{2}+1%
\right] r_{+}^{8p+5}}\Upsilon \ .  \label{dMSS1}
\end{equation}
where
\begin{eqnarray}
\Upsilon &=&2^{2p}l^{4}\Xi ^{2}q^{4(p+1)}+\frac{2^{p+1}}{(2p+1)}%
q^{2(p+1)}l^{2}r_{+}^{4\left( p+1\right) }[2(p+1)-\Xi ^{2}]+ \\
&&[4(p+1)\left( \Xi ^{2}-1\right) +\Xi ^{2}]r_{+}^{8\left( p+1\right) }
\nonumber
\end{eqnarray}
\begin{figure}[tbp]
\epsfxsize=10cm \centerline{\epsffile{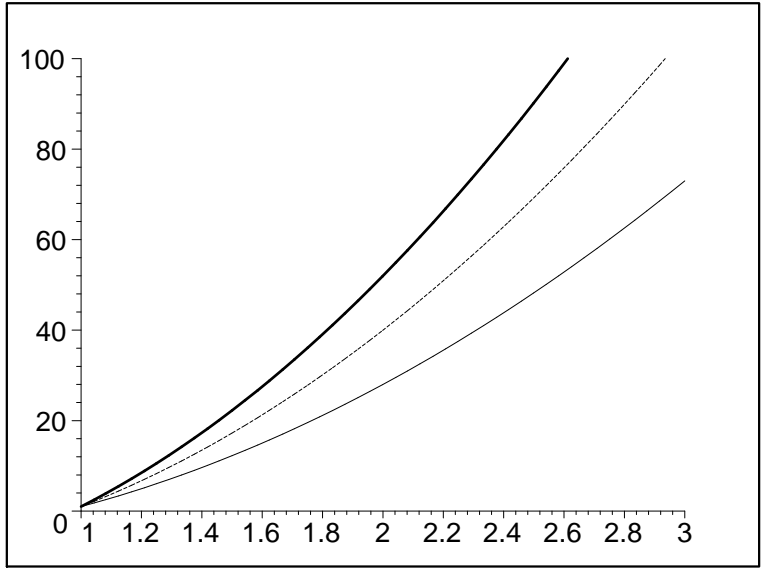}}
\caption{$\Upsilon $ versus $\Xi $ for $r_{+}=1$, $l=0.1$, $q=0.1$ and $p=1$
(continuous line), $p=2$ (dotted line) and $p=3$ (bold line).}
\label{Fig1}
\end{figure}
The heat capacity is positive for $m\geq m_{\mathrm{ext}}$, where the
temperature is positive. This fact can be seen easily for $\Xi =1$, where
the $\Upsilon $ is positive and then $(\partial ^{2}M/\partial S^{2})_{%
\mathbf{J},Q}$ is positive too. Also, one may see from Fig. \ref{Fig1} that
the $\Upsilon $ increases as $\Xi $ increases, and therefore it is always
positive. Thus, the black brane is stable in the canonical ensemble. In the
grand-canonical ensemble, after some algebraic manipulation, we obtain
\begin{equation}
\mathbf{H}_{S,\mathbf{J},Q}^{M}=\frac{64\pi \left[ \left( 4p+3\right)
r_{+}^{4\left( p+1\right) }+\left( 2p+1\right) 2^{p}l^{2}q^{2\left(
p+1\right) }\right] \left[ r_{+}^{4\left( p+1\right) }+2^{p}l^{2}q^{2\left(
p+1\right) }\right] ^{-1}}{\left[ (4p+1)\Xi ^{2}+1\right] \left[ \left(
4p+3\right) \left( 2p+1\right) (p+1)2^{p}l^{2}\Xi ^{6}r_{+}^{8p+5}q^{2p}%
\right] }.  \label{dHes}
\end{equation}
It is easy to see $\mathbf{H}_{S,\mathbf{J},Q}^{M}$ is positive for all the
allowed values of $q\leq q_{\mathrm{ext}}$. Thus, the $(4p+4)$-dimensional
asymptotically AdS charged rotating black brane is locally stable in the
grand-canonical ensemble.

\section{ CLOSING REMARKS}

First of all, we considered Einstein gravity coupled to a conformally
invariant Maxwell field. In $d$-dimensions the action of Maxwell field is
shown to enjoy the conformal invariance if the power is chosen as $d/4$.
Then, For obtaining finite action, we combined well-defined
Einstein-conformally invariant Maxwell action with conterterm action for
which, the last action arise from AdS/CFT correspondence. After that, we
found a class of charge rotating solutions in Einstein-conformally invariant
Maxwell gravity in the presence of a cosmological constant. These solutions
may be interpreted as black brane solutions with inner and outer event
horizons or an extreme black brane when $m>m_{\mathrm{ext}}$ or $m=m_{%
\mathrm{ext}}$, respectively. The black holes presented here differ from the
standard higher-dimensional solutions \cite{Rastegar} since (a) the scalar
curvature of the spacetimes is only related to cosmological constant, and
(b) the electric charge term in the metric coefficient goes as $%
r^{-(4p+2)}$ and in the standard case is $r^{-2(4p+1)}$.

Also, we computed physical quantities of the black brane solutions such as
the temperature, the angular velocity, the electric charge and the
potential, and showed that the expression of the potential does not depend
on the dimension. Also, we obtained a Smarr-type formula and found that for
these black brane solutions, these conserved and thermodynamic quantities
satisfy the first law of thermodynamics. Finally, we studied the phase
behavior of the rotating black branes and showed that there is no
Hawking--Page phase transition in spite of conformally invariant Maxwell
field. Indeed, we investigated stability of black brane in both the
canonical and grand-canonical ensembles and show that the system is stable
in the whole phase space.

\begin{acknowledgements}
We thank Professor M. H. Dehghani for useful discussions and
enlightening comments. This work has been supported financially by
Research Institute for Astronomy and Astrophysics of Maragha.
\end{acknowledgements}


\vspace{0.5cm}

\begin{thebibliography}{99}
\bibitem{Mal} J. Maldacena, Adv. Theor. Math. Phys. 2 (1998) 231;\newline
E. Witten, Adv. Theor. Math. Phys. 2 (1998) 253;\newline S.W.
Hawking, C.J. Hunter, M.M. Taylor-Robinson, Phys. Rev. D 59 (1999)
064005;\newline
O. Aharony, S.S. Gubser, J. Maldacena, H. Ooguri,
Y. Oz, Phys. Rep. 323 (2000) 183.

\bibitem{Sken} M. Hennigson, K. Skenderis, J. High Energy Phys. 7 (1998)
023;\newline
 S. Nojiri, S.D. Odintsov, Phys. Lett. B 444 (1998)
92;\newline
 V. Balasubramanian, P. Kraus, Commun. Math. Phys. 208
(1999) 413;
\newline S. Nojiri, S.D. Odintsov, S. Ogushi, Phys. Rev. D 62
(2000) 124002; \newline R.G. Cai, Phys. Rev. D 63 (2001) 124018;
\newline M.H. Dehghani, Phys. Rev. D 65 (2002) 124002; \newline
M.H. Dehghani, Phys. Rev. D 66 (2002) 044006; \newline A.M.
Ghezelbash, D. Ida, R.B. Mann, T. Shiromizu, Phys. Lett. B 535
(2002) 315.

\bibitem{Bailin} D. Bailin, A. Love, Rep. Prog. Phys. 50 (1987) 1087.

\bibitem{Nieu} P.V. Nieuwenhuizen, Phys. Rep. 68 (1981) 189.

\bibitem{Freu} P.G.O. Freund, M. Rubin, Phys. Lett. B 97 (1980) 233.

\bibitem{comment} Consider the Lagrangian of the form $L(F)$, where $%
F=F_{\mu \nu}F^{\mu \nu}$. It is easy to show that for this form of
Lagrangian in $d$-dimensions, $T_{\mu}^{\mu}=4\left[F\frac{dL}{dF}-\frac{d}{4%
}L\right]$; so $T_{\mu}^{\mu}=0$ implies $L(F)=Constant\times F^{\frac{d}{4}%
} $.

\bibitem{Haissene} M. Hassaine, J. Math. Phys. (N.Y.) 47 (2006) 033101.

\bibitem{AllStaticEM} T. Thiemann, Nucl. Phys. B 436 (1995) 681; \newline
M.H. Dehghani, A. Khodam-Mohammadi, Phys. Rev. D 67 (2003) 084006; \newline
D. Horvat, S. llijic, Z. Narancic, Class. Quantum Grav. 22 (2005) 3817;
\newline
M. Hassaine, C. Martinez, Phys. Rev. D 75 (2007) 027502.

\bibitem{AllRotBH} G. Clement, Phys. Rev. D 57 (1998) 4885; \newline
A.M. Awad, C.V. Johnson, Phys. Rev. D 63 (2001) 124023; \newline
D. Ida, Y. Uchida, Phys. Rev. D 68 (2003) 104014; \newline
G.W. Gibbons, H. Lu, D.N. Page, C.N. Pope, Phys. Rev. Lett. 93 (2004)
171102; \newline
M. Cvetic, H. Lu, D.N. Page, C.N. Pope, Phys. Rev. Lett. 95 (2005) 071101;
\newline
G.W. Gibbons, M.J. Perry, C.N. Pope, Class. Quantum Grav. 22 (2005) 1503;
\newline
B.M.N. Carter, I.P. Neupane, Phys. Rev. D 72 (2005) 043534; \newline
A.N. Aliev, Phys. Rev. D 74 (2006) 024011; \newline
H. Ishihara, M. Kimura, K. Matsuno, S. Tomizawa, Phys. Rev. D 74 (2006)
047501; \newline
J. Kunz, F. Navarro-Lerida, E. Radu, gr-qc/0702086; \newline
Y. Brihaye, T. Delsate, gr-qc/0703146; \newline
M.H Dehghani, S.H. Hendi, A. Sheykhi, H. Rastegar Sedehi, JCAP 0702 (2007)
020.

\bibitem{dehhend2} M.H. Dehghani, S.H. Hendi, Int. J. Mod. Phys. D 16 (2007)
1829.

\bibitem{Rastegar} M.H. Dehghani, H.R. Sedehi, Phys. Rev. D 74 (2006) 124018.

\bibitem{Hawpag} S.W. Hawking, D.N. Page, Commun. Math. Phys. 87 (1983) 577;
\newline
M. Cvetic, S. Nojiri, S.D. Odintsov, Nucl. Phys. B 628 (2002) 295.

\bibitem{Birmingham} D. Birmingham, Class. Quantum Grav. 16 (1999) 1197.

\bibitem{BCCC} D.R. Brill, J. Louko, P. Peldan, Phys. Rev. D 56 (1997) 3600;
\newline
R.G. Cai, K.S. Soh, Phys. Rev. D 59 (1999) 044013; \newline
A. Chamblin, R. Emparan, C.V. Johnson, R.C. Myers, Phys. Rev. D 60 (1999)
064018; \newline
R.G. Cai, A.Z. Wang, Phys. Rev. D 70 (2004) 064013.

\bibitem{Myers} R.C. Myers, Phys. Rev. D 36 (1987) 392.

\bibitem{Brown} J.D. Brown, J.W. York, Phys. Rev. D 47 (1993) 1407.

\bibitem{LemAw} J.P.S. Lemos, V.T. Zanchin, Phys. Rev. D 54 (1996) 3840;
\newline
A.M. Awad, Class. Quantum Grav. 20 (2003) 2827.

\bibitem{Haw1} S.W. Hawking, Commun. Math. Phys. 25 (1972) 152; \newline
S.W. Hawking, G.F.R. Ellis, \textit{The Large Scale of SpaceTime},
(Cambridge University Press, 1973).

\bibitem{dehbordshah} M.H. Dehghani, G.H. Bordbar, M. Shamirzaie, Phys. Rev.
D 74 (2006) 064023.

\bibitem{Beck} J.D. Beckenstein, Phys. Rev. D 7 (1973) 2333; \newline
S.W. Hawking, C.J. Hunter, Phys. Rev. D 59 (1999) 044025; \newline
C.J. Hunter, Phys. Rev. D 59 (1999) 024009; \newline S.W. Hawking,
C.J. Hunter, D.N. Page, Phys. Rev. D 59 (1999) 044033; \newline
R.B. Mann, Phys. Rev. D 60 (1999) 104047; \newline R.B. Mann,
Phys. Rev. D 61 (2000) 084013.

\bibitem{Gub} M. Cvetic, S.S. Gubser, J. High Energy Phys. 04 (1999) 024;
\newline
M.M. Caldarelli, G. Cognola, D. Klemm, Class. Quantum Grav. 17 (2000) 399.
\end{thebibliography}
\end{document}